# Structural and Electronic Properties of Graphdiyne Carbon Nanotubes from Large-Scale DFT Calculations

*Sangavi Pari,[1] Abigail Cuéllar,[2] and Bryan M. Wong[1]\**

[1]Department of Chemical & Environmental Engineering and Materials Science & Engineering Program

University of California-Riverside, Riverside, CA 92521, USA

[2]Rialto High School, Rialto, CA 92376, USA

\*Corresponding author. E-mail: bryan.wong@ucr.edu. Web: http://www.bmwong-group.com

**Abstract.** Using large-scale DFT calculations, we have investigated the structural and electronic properties of both armchair and zigzag graphdiyne nanotubes as a function of size. To provide insight in these properties, we present new detailed calculations of the structural relaxation energy, effective electron/hole mass, and size-scaling of the bandgap as a function of size and chirality using accurate screened-exchange DFT calculations. These calculations provide a systematic evaluation of the structural and electronic properties of the largest graphdiyne nanotubes to date – up to 1,296 atoms and 23,328 basis functions. Our calculations find that zigzag graphdiyne nanotubes (GDNTs) are structurally more stable compared to armchair GDNTs of the same size. Furthermore, these large-scale calculations allow us to present simple analytical formulae to guide future experimental efforts for estimating the fundamental bandgaps of these unique nanotubes as a function of chirality and diameter. While the bandgaps for both the armchair and zigzag GDNTs can be tuned as a function of size, the conductivity in each of these two different chiralities is markedly different. Zigzag GDNTs have wider valence and conduction bands and are expected to have a higher electron- and hole-mobility than their armchair counterparts.

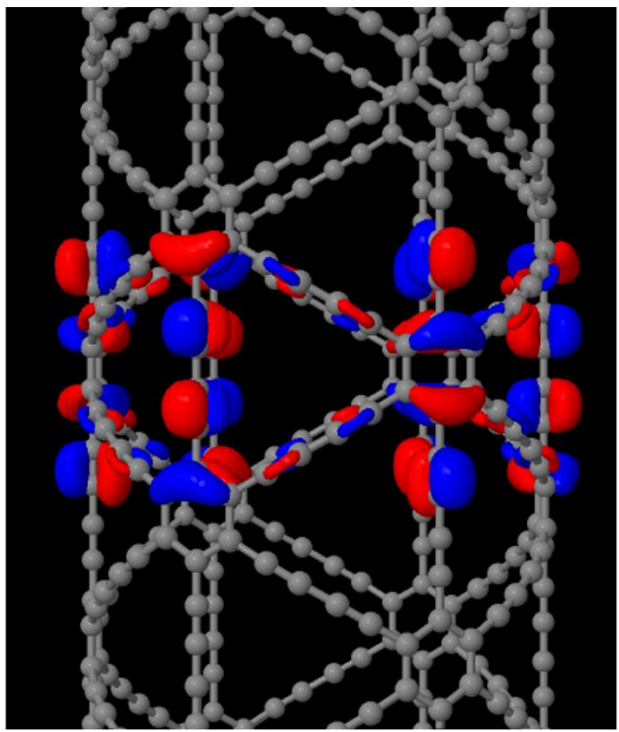 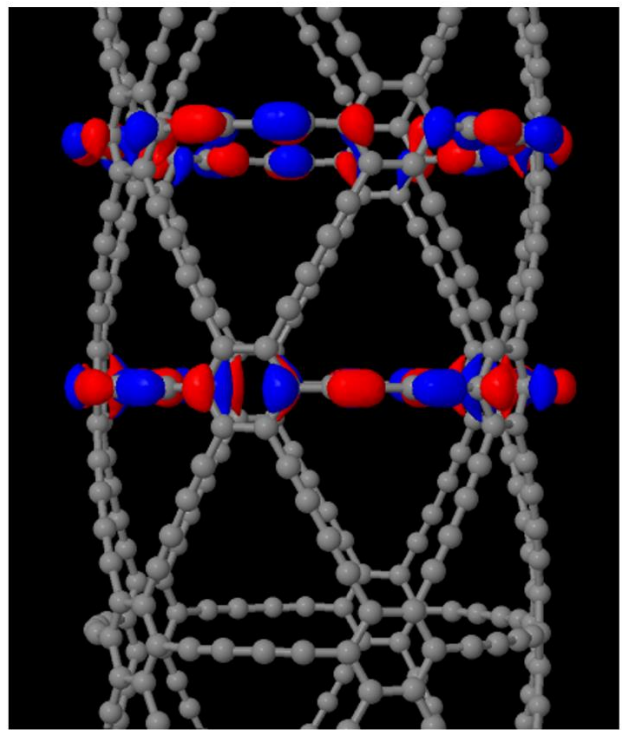

Table of Contents Figure

**I. Introduction**

Carbon nanotubes and related allotropes continue to garner immense interest due to the unique electronic properties that naturally arise from their intrinsic one-dimensional nature.[1] Specifically, one-dimensional nanosystems (such as nanowires and nanotubes) are the smallest dimensions that can be used for efficient transport of electrons and are, therefore, critical to the functionality of nanoscale devices.[2] Within the carbon nanotube family, these devices have included field effect transistors,[3-8] actuators,[9, 10] nanotube films for flexible displays,[11] and nanotube hybrid solar panels.[12] In many of these devices, carbon nanotubes of a specific chirality (or a narrow range of chiralities possessing similar electronic properties) are often required. Within a normal distribution of carbon nanotube chiralities, roughly one-third of nanotubes are metallic while the other two-thirds exhibit semi-conducting behavior.[13] Because of this wide variation in carbon nanotube chiralities, recent efforts have focused on other allotropes of carbon to achieve detailed control over their electronic properties and device functionality.

In recent years, much effort has focused on graphdiyne (cf. Fig. 1), which is a new allotrope of carbon composed of two acetylenic linkages (with *sp*-hybridized carbon atoms) between nearest-neighbor hexagonal rings (composed of $sp^2$-hybridized carbons). Planar graphdiyne exhibits a high-temperature stability and semi-conducting properties comparable to silicon[14] and has been proposed for gas separation applications,[15] nanoscale devices,[16] photocatalysts for hydrogen production,[17] and hydrogen purification in syngas production.[18] In a relatively recent report,[19] the experimental synthesis and construction of graphdiyne nanotubes (GDNTs, see Fig. 2) were carried out for the very first time, and subsequent papers on other graphdiyne-based nanostructures have reported unique electronic properties, including charge mobilities as high as $2 \times 10^5$ cm$^2$ V$^{-1}$ s$^{-1}$ at room temperature.[20] However, to the best of our knowledge, a systematic study on the structural and electronic properties of GDNTs as a function of size and chirality has not been previously reported. To provide insight in these properties, we present a new, detailed investigation of the structural relaxation energy, effective electron/hole mass, and size-scaling of the bandgap as a function of size and chirality using accurate screened-exchange DFT calculations. These calculations provide a systematic evaluation of the structural and electronic properties of the largest graphdiyne nanotubes to date – up to 1,296 atoms and 23,328 basis functions. Furthermore these large-scale calculations allow us to present simple analytical formulae to guide future experimental efforts for estimating the fundamental bandgaps of these unique nanotubes as a function of chirality and diameter as well as provide a detailed understanding of the size-scaling of structural and electronic properties. Finally, we give a detailed analysis of all these effects for both the armchair and zigzag GDNTs and discuss the implications of these computed properties on electron/hole mobility and potential applications of these results.

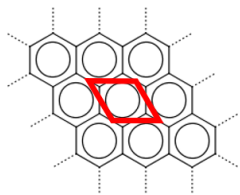

**graphene**

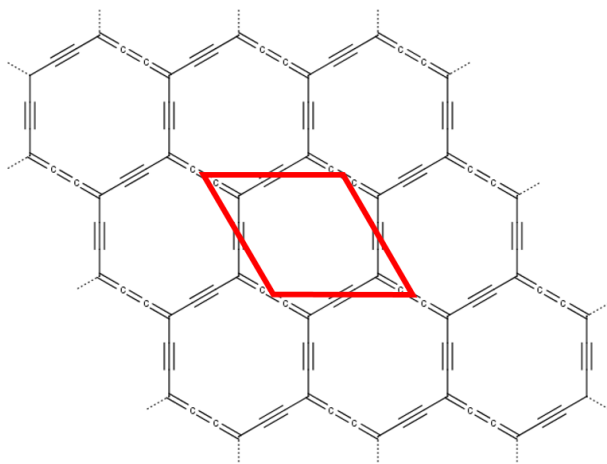

**α-graphyne**

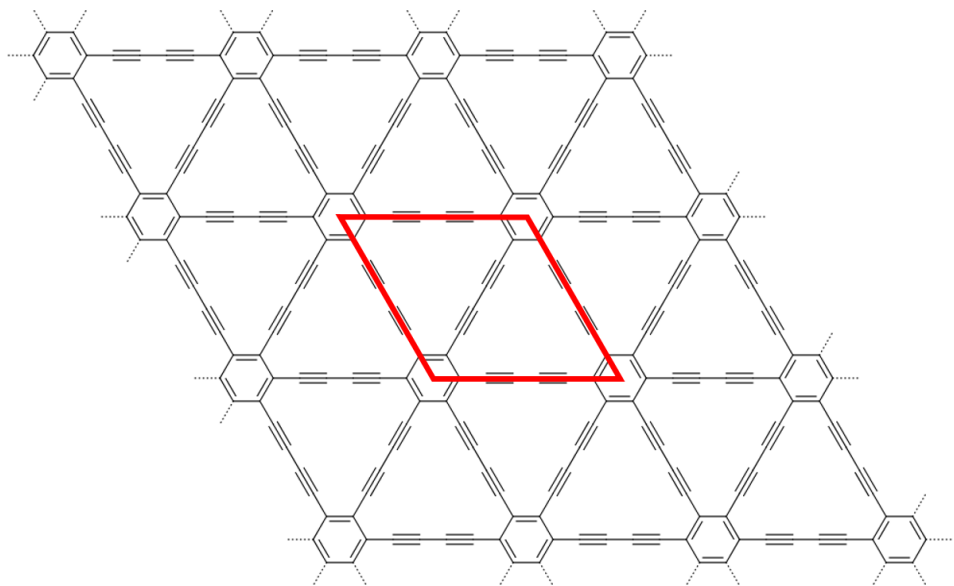

**graphdiyne**

**Figure 1:** Chemical structures and unit cells for graphene, α-graphyne, and graphdiyne. All structures and unit cells are drawn to scale with each unit cell containing 2, 8, and 18 carbon atoms for graphene, α-graphyne, and graphdiyne, respectively.

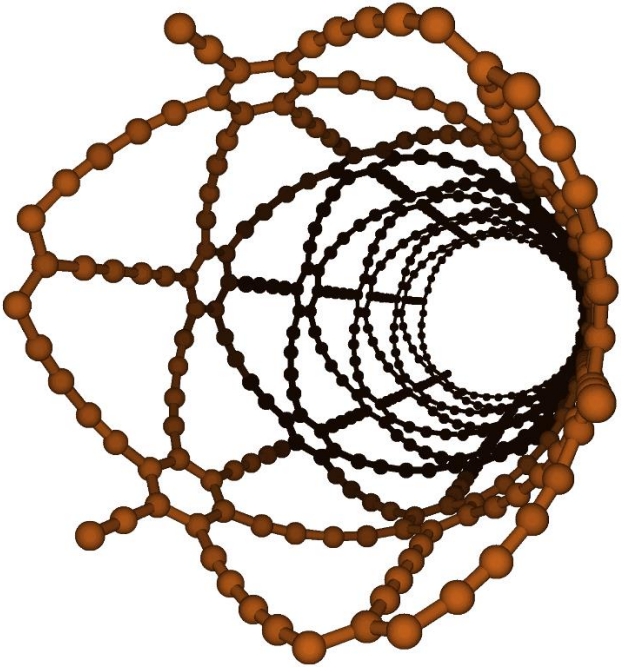 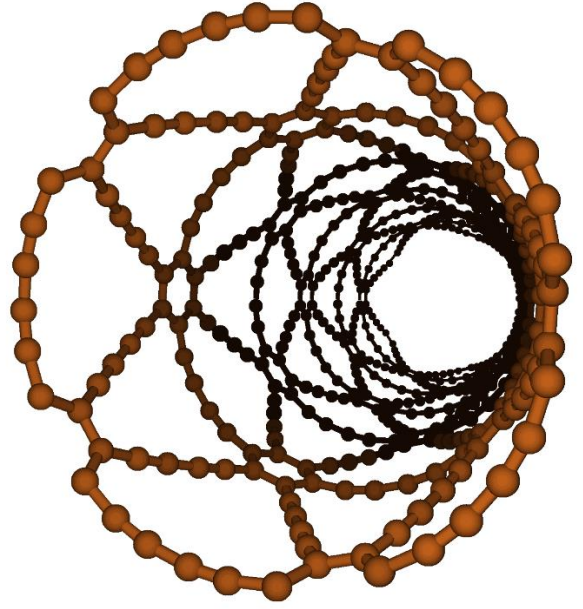

(3,3) Zigzag GDNT     (5,0) Armchair GDNT

**Figure 2:** Optimized structures of the (3,3) zigzag and (5,0) armchair graphdiyne nanotubes.

## II. Structural Properties of Graphdiyne Nanotubes

Planar graphdiyne belongs to the *p6mm* space group, and its unit cell is defined by the two lattice vectors $\vec{a}_1 = a\hat{x}$ and $\vec{a}_2 = \frac{a}{2}(-\hat{x} + \sqrt{3}y)$, as shown in Fig. 3. Any GDNT of arbitrary chirality can be generated by these two vectors through the chiral vector $\vec{C}_h = n\vec{a}_1 - m\vec{a}_2$, where $|\vec{C}_h| = ac_h$, $[c_h \equiv (n^2 + m^2 + nm)^{1/2}]$, and the tube diameter is given by $d_t = \frac{ac_h}{\pi}$. The chiral angle as shown in Fig. 3 is defined by $\cos\theta = \frac{2n+m}{2c_h}$, where $0 \leq \theta \leq \pi/6$. Based on these definitions for the chiral vector and chiral angle, armchair GDNTs ($\theta = 0$) are represented by the (*n*, 0) chiral index, and zigzag GDNTs ($\theta = \pi/6$) are characterized by the (*n*, *n*) chiral index, *which is the opposite convention in carbon*

*nanotubes*. Fig. 3 illustrates the lattice vectors and selected examples of chiral vectors for a (3, 0) armchair and (2, 2) zigzag GDNT.

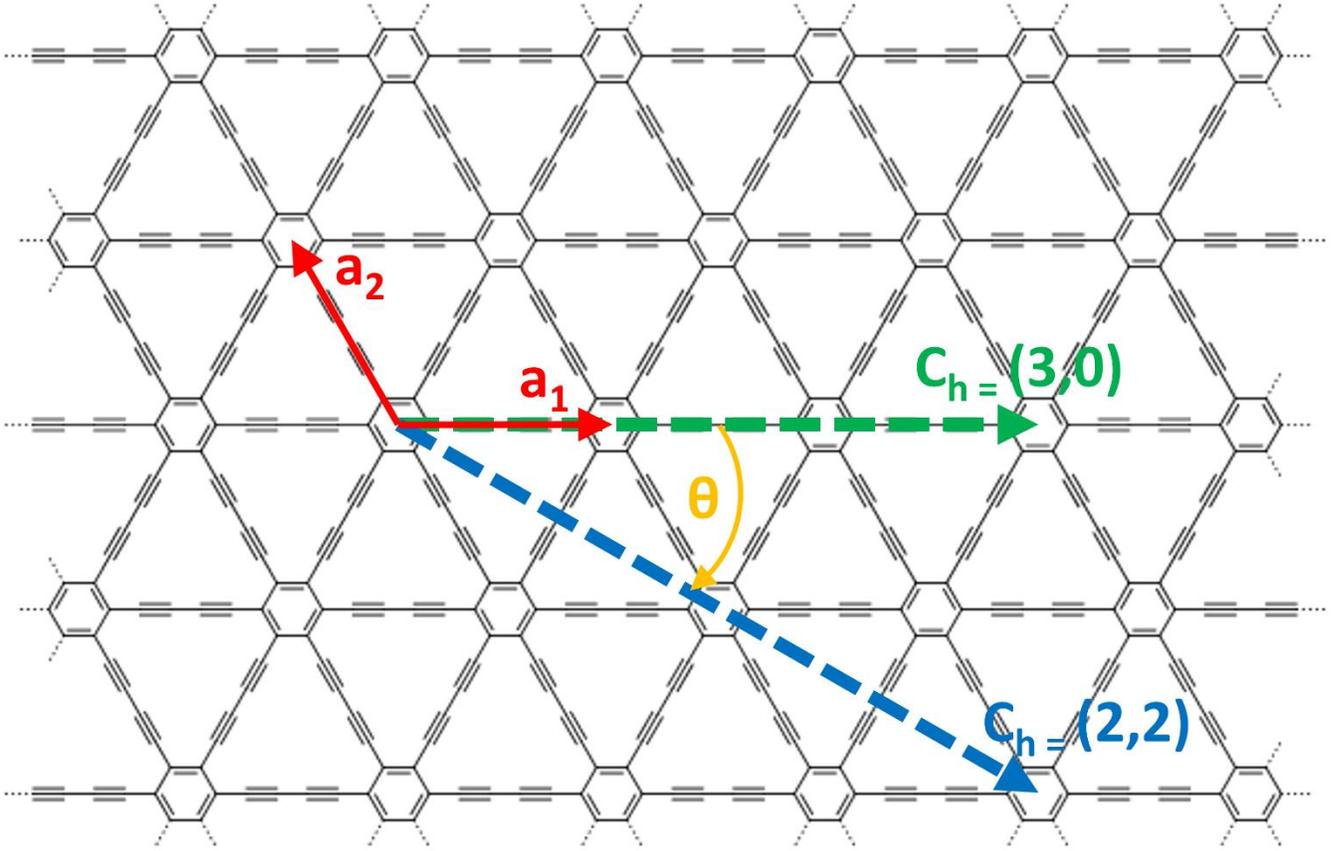

**Figure 3:** Lattice vectors $\vec{a}_1$ and $\vec{a}_2$, chiral angle $\theta$, and selected chiral vectors (3,0) and (2,2) for a graphydiyne sheet.

### III. Theory and Methodology

All calculations were carried out with a massively-parallelized version of the CRYSTAL14 program,[21] which has the capability of using both all-electron Gaussian-type orbitals and exact Hartree-Fock exchange within periodic boundary conditions. The latter is particularly important for obtaining accurate electronic properties for periodic systems since the incorporation of Hartree-Fock exchange can partially correct for electron-delocalization errors inherent to both LDA (local density approximation) and GGA (generalized gradient approximation) exchange-correlation functionals. For this reason, we utilized the range-separated HSE06 functional[22] for obtaining the electronic properties for both the graphdiyne sheet and all of the zigzag and armchair graphdiyne nanotubes. As opposed to range-separated hybrid DFT methods that incorporate a "full" range separation of 100% asymptotic Hartree-Fock exchange (typically used in isolated molecules[23-28]), the HSE06 functional incorporates a screened Hartree-Fock exchange that decays to zero. Most importantly, our previous studies[29] with the HSE06 functional in the CRYSTLA14 program has shown that it much more computationally efficient than conventional global hybrid functionals[30] and is significantly more accurate than conventional semi-local functionals. It is worth noting that although the HSE06 calculations are more efficient than conventional hybrid DFT methods, the calculations on some of the largest GDNTs were still extremely computationally intensive due to the immense size of these nanotubes. For example, the largest of these structures (specifically the (36,0) armchair GDNT), consists of 1,296 atoms and 23,328 basis functions and, as such, this study constitutes the largest systematic study of these nanostructures to date.

Geometries for all of the graphydiyne nanotubes were optimized using a large TZVP all-electron basis set[31] with one-dimensional periodic boundary conditions along the tube axis. At the optimized geometries, a final single-point HSE06 calculation was performed with 100 k points along the one-dimensional Brillouin zone to obtain the electronic band structure for all of the nanotube geometries.

**IV. Results and Discussion**

**A. Benchmark Calculations**

Since a systematic study of the electronic properties of armchair and zigzag GDNTs has not been previously investigated, we first benchmarked our HSE06/TZVP results for the graphdiyne sheet against the high-level $G_0W_0$ (Green's function G and screened Coulomb interaction W) calculations by Luo *et al*.[32] In this previous study, the $G_0W_0$ bandgap of the planar graphdiyne sheet attains a value of 1.10 eV, which is consistent with experimental measurements of graphdiyne film. The band structure along high-symmetry points in the graphdiyne Brillouin zone (defined by the high-symmetry points Γ, X, and M in momentum space) obtained by our HSE06/TZVP calculations is shown in Fig. 4. We obtain a direct bandgap of 1.26 eV at the Γ point, which is in relatively good agreement with the computationally-intensive $G_0W_0$ bandgap of 1.10 eV. The close agreement between our HSE06 results is in stark contrast to conventional PBE calculations which severely underestimate the bandgap by more than 50%, giving a value of 0.54 eV. As such, our benchmark calculations for the bandgap of planar graphdiyne demonstrate that our HSE06 calculations are reasonable for our parametric studies on the various GDNTs studied in this work.

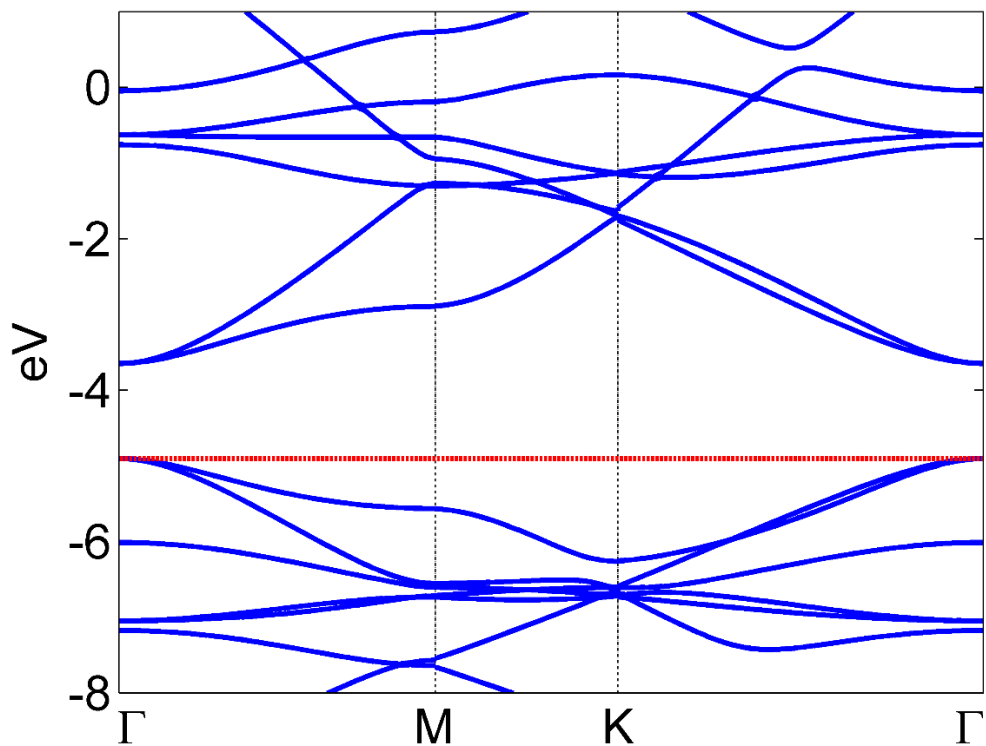

**Figure 4.** Electronic band structure for planar graphdiyne obtained at the HSE06/TZVP level of theory. The dashed horizontal line indicates the position of the Fermi energy, and a direct bandgap of 1.26 eV occurs at the Γ point within the irreducible Brillouin zone.

## B. Structural Properties

To give deeper insight into the structural stability of all the nanotubes relative to the planar graphdiyne sheet, we calculated the relaxation energy, $\Delta E$, given by

$$\Delta E = E_{\text{nanotube}} - n \cdot E_{\text{sheet}}, \tag{1}$$

where $E_{\text{nanotube}}$ is the electronic energy of the geometry-optimized nanotube, $E_{\text{sheet}}$ is the electronic energy of the graphdiyne sheet, and $n$ is the number of repeat units along the nanotube circumference (which also corresponds to the first chiral index for each $(n,m)$ nanotube). Fig. 5 shows that $\Delta E$ decreases monotonically with size, and the stability becomes comparable to planar graphdiyne for GDNT diameters larger than 9 nm. To further test the structural stability of these nanostructures, we calculated the harmonic frequencies for the smallest (2,2) GDNT, which contains 72 atoms in its primitive unit cell (harmonic frequency calculations for other larger GDNTs were computationally out of reach due to their immense size; for example, the largest GDNT in this work contains up to 1,296 atoms and 23,328 basis functions). At the optimized geometry, we obtained real-valued frequencies for all of the vibrational modes for the (2,2) GDNT (vibrational frequencies, symmetries, and infrared/Raman analysis for the (2,2) GDNT can be found in the Supporting Information) . Most importantly, since the (2,2) GDNT is the most strained nanotube in this study (cf. Fig. 5), our stability analysis also implies that the other larger, less-strained GDNTs are also structurally stable. We also tabulated the binding energy per atom for all GDNTs in Tables 1 and 2 using the expression

$$E_{\text{binding}} = \frac{1}{N}(E_{\text{nanotube}} - N \cdot E_{\text{atom}}), \tag{2}$$

where $N$ is the number of atoms in the nanotube, $E_{\text{nanotube}}$ is the electronic energy of the geometry-optimized nanotube, and $E_{\text{atom}}$ is the total atomic energy of the carbon atom (in its ground triplet state). Similar to the computed relaxation energies, the binding energy per atom decreases monotonically with diameter and becomes nearly constant for GDNT diameters larger than 9 nm. To compare the structural stabilities of these GDNTs against conventional nanostructures, we also calculated the binding energy per atom for a conventional (13,0) carbon nanotube which has a similar diameter to a (2,2) zigzag GDNT. At the HSE06/TZVP level of theory, we obtain a binding energy per atom of -7.9480 eV, which is 0.88 eV more stable than a similarly-sized (2,2) zigzag GDNT. It is also interesting to note that the zigzag GDNTs are structurally more stable compared to armchair GDNTs of the same size. This trend can be rationalized since the geometric structures of the armchair and zigzag GDNTs are topologically different. Specifically, all six of the acetylenic linkages (between the benzene rings) in armchair GDNTs straddle the circumference of the nanotube, whereas two of the acetylenic linkages in the zigzag GDNTs are oriented along the nanotube axis (cf. Fig. 2), which partially relieves these strain effects around the GDNT circumference.

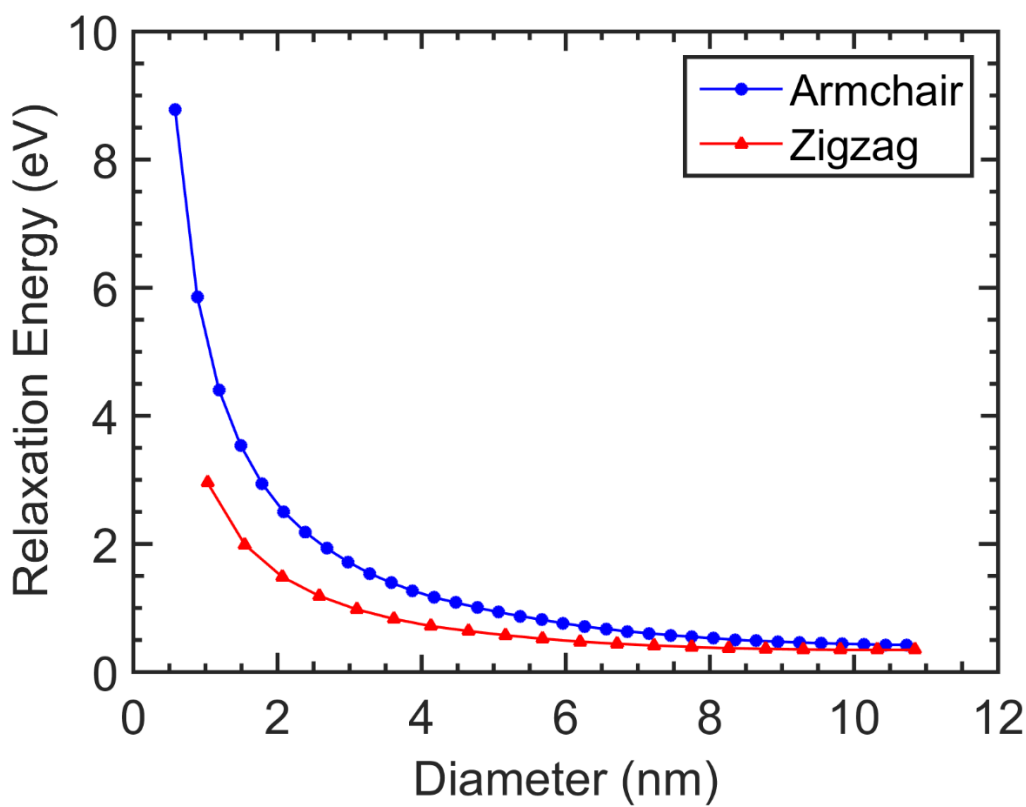

**Figure 5.** Relaxation energy, $\Delta E$, as a function of diameter for both armchair and zigzag GDNTs obtained at the HSE06/TZVP level of theory.

## C. Electronic Properties

Figs. 6 and 7 plot the band structures of selected armchair and zigzag GDNTs, respectively, along the irreducible Brillouin zone (defined by the high-symmetry points Γ and X in momentum space). In all of the different chiralities, we find that the electronic band structures are characterized by a direct bandgap at the Γ point. We calculated the effective mass $m^*$ of the electrons and holes at the conduction band minimum and valence band maximum, respectively, using the expression

$$m = \pm \hbar^2 \left(\frac{d^2 E}{dk^2}\right)^{-1}. \tag{3}$$

The positive sign is taken for the (electron) conduction band, and the negative sign corresponds to the (hole) valence band. Tables 1 and 2 give a summary of the various structural and electronic properties (radii, relaxation energies, binding energy per atom, effective electron mass, effective hole mass and bandgaps) of the armchair and zigzag GDNTs examined in this study.

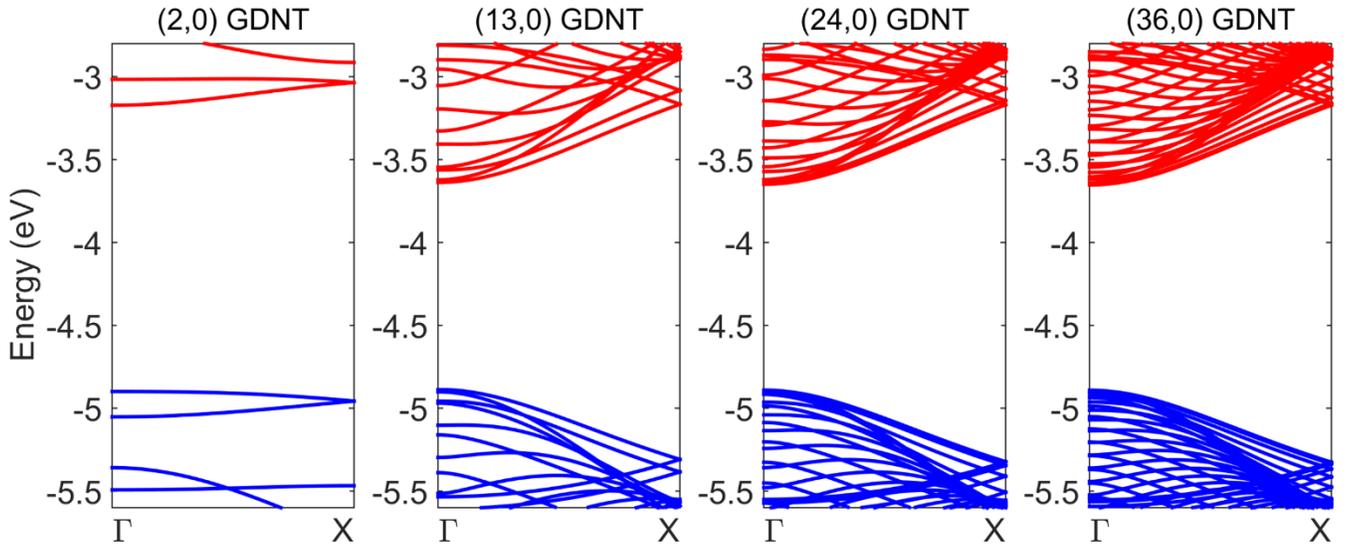

**Figure 6.** Electronic band structures (relative to vacuum at 0 eV) of various (*n*,0) armchair GDNTs for *n* = 2, 13, 24, and 36. Note the narrow (and nearly dispersionless) bands for the (2,0) GDNT.

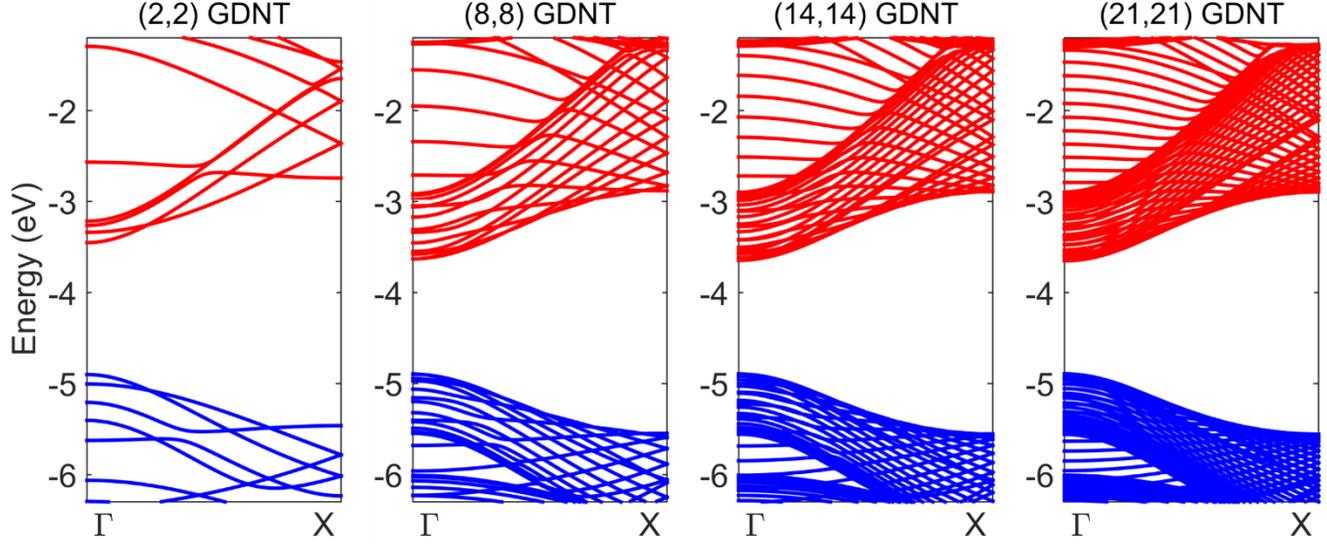

**Figure 7.** Electronic band structures (relative to vacuum at 0 eV) of various (*n,n*) zigzag GDNTs for *n* = 2, 8, 14, and 21. Note that the zigzag GDNTs have wider valence and conduction bands compared to their armchair GDNT counterparts.

Fig. 8 plots the bandgap of the armchair and zigzag GDNTs as a function of nanotube radius. Using our HSE06/TZVP calculations, we performed a nonlinear fit of the bandgap ($E_g$) as a function of diameter (*d*). We chose a flexible functional form given by $E_g = A/d + B$, where *A* and *B* are independent free parameters subject to our nonlinear least-squares fit. Based on our HSE06 bandgaps, we obtained fitted expressions

$$E_g(\text{armchair}) = \frac{0.24 \text{ eV}}{d \text{ (in nm)}} + 1.2 \text{ eV}, \tag{4}$$

$$E_g(\text{zigzag}) = \frac{0.21 \text{ eV}}{d \text{ (in nm)}} + 1.2 \text{ eV}, \tag{5}$$

with *R*-squared fit values of 0.87 and 0.97, respectively (the slightly lower *R*-squared fit value for the armchair GDNTs arises from larger strain values compared to their zigzag GDNTs counterparts [cf. Fig. 5]). It is interesting to note that the last constant term in Eqs. (4) and (5) corresponds closely to the bandgap of the planar graphdiyne sheet; in other words, the constant term in Eqs. (4) and (5) yields the bandgap of a GDNT having an infinite diameter. Although we determined this constant as a free parameter in our fit, it is noteworthy to point out that we nearly recover the bandgap of planar graphdiyne calculated earlier in

Section IV.A (we do not obtain the exact bandgap of planar graphdiyne due to relatively strong curvature effects that are still present in the larger GDNTs).

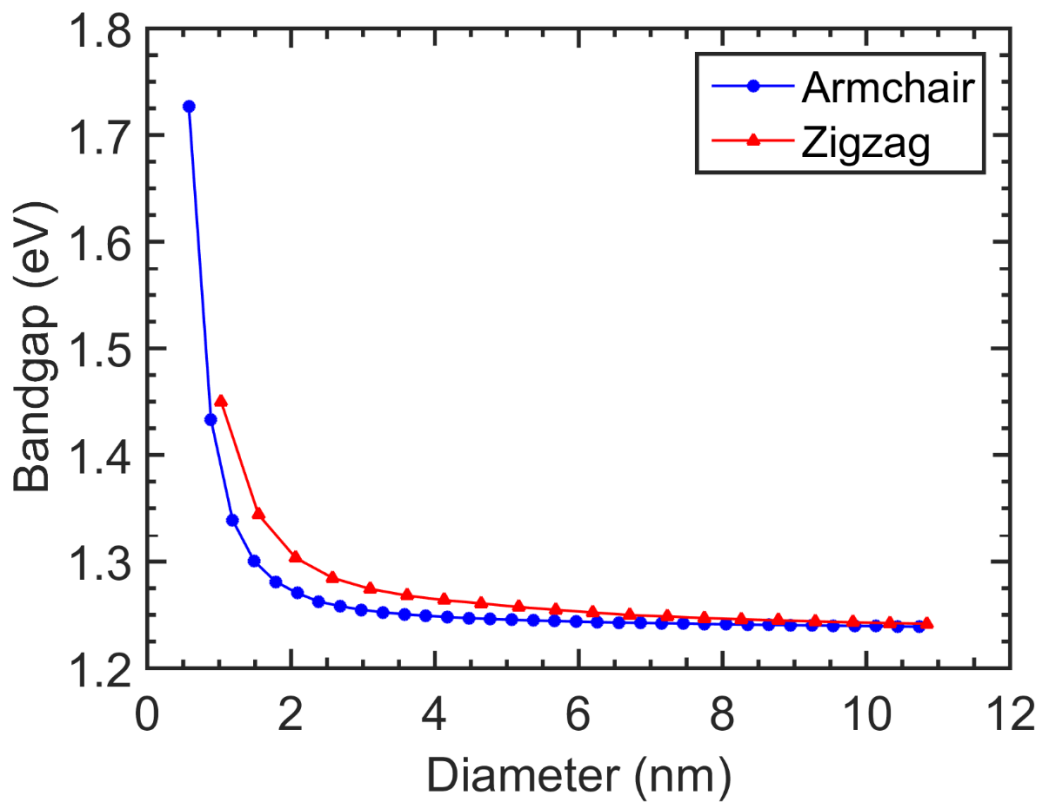

**Figure 8.** Electronic bandgap as a function of diameter for both armchair and zigzag GDNTs obtained at the HSE06/TZVP level of theory.

Finally, we examine in greater detail the electronic band structures of both the armchair and zigzag GDNTs. As shown in Figs. 6 and 7, the armchair GDNTs possess narrower valence and conduction bands, whereas the zigzag GDNTs exhibit much wider bands (band structures for all 35 armchair and all 20 zigzag GDNTs can be found in the Supporting Information). Specifically, the width of an electronic band reflects the orbital interactions along the nanotube axis, with wide bands denoting orbital delocalization and narrow bands corresponding to localization (small overlap). To corroborate these findings, we plotted the highest occupied crystal orbitals (HOCO) and lowest unoccupied crystal orbitals (LUCO) at the Γ point for both the armchair and zigzag GDNTs (using the same isosurface values for each). Fig. 9 shows that both the HOCO and LUCO in armchair GDNTs are localized on the acetylenic linkages along the circumference of the nanotube. In contrast, for zigzag GDNTs, the HOCO and LUCO are delocalized along the entire axis of zigzag and, therefore, both hole- and electron-transport are more facile in zigzag GDNTs compared to their armchair counterparts. While the bandgaps for both the armchair and zigzag GDNTs can certainly be tuned as a function of size, the conductivity in each of these two different chiralities is markedly different. Zigzag GDNTs have wider valence and conduction bands (which are demonstrated by the orbital diagrams in Fig. 9 and the projected density of states plots in Fig. 10) and are, therefore, expected to have a higher conductivity than their armchair counterparts. As such, both the armchair and zigzag chiralities provide an additional intrinsic material property that can be used to modulate both hole- and electron-transport in photo-induced applications and processes.

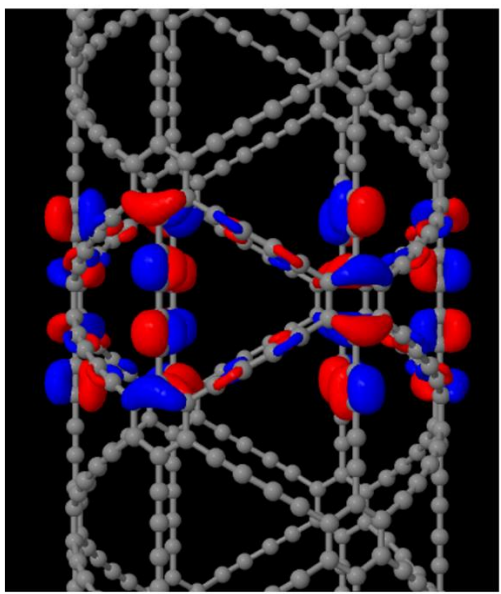 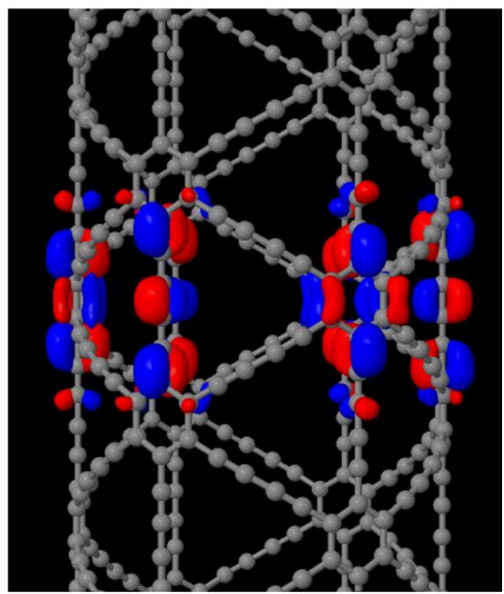

(3,3) GDNT HOCO      (3,3) GDNT LUCO

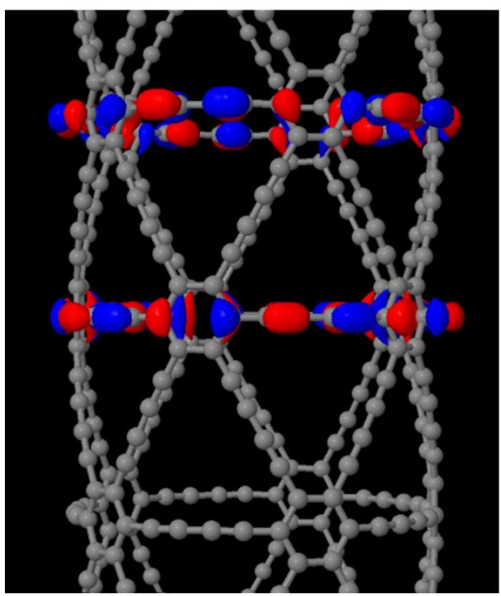 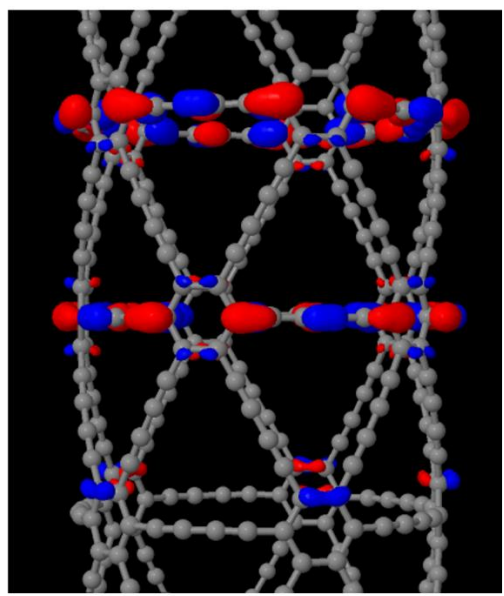

(5,0) GDNT HOCO      (5,0) GDNT LUCO

**Figure 9.** Highest occupied and lowest unoccupied crystal orbitals (HOCO and LUCO) for the (5,0) armchair and (3,3) zigzag GDNTs (only crystal orbitals within one unit cell are shown for clarity). Both the HOCO and LUCO are localized along acetylenic linkages along the circumference of the (5,0) GDNT, whereas the HOCO and LUCO are localized along acetylenic linkages along the axis of the (3,3) GDNT.

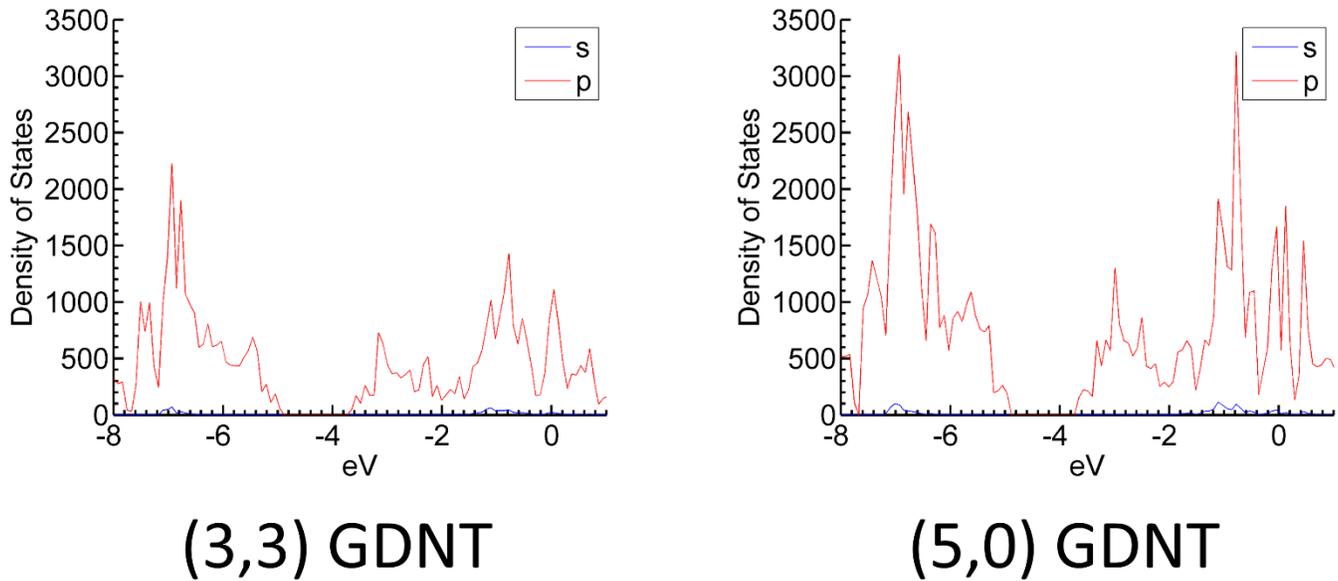

**Figure 10.** Projected density of states for the (3,3) zigzag and (5,0) armchair GDNT. For both the (3,3) and (5,0) GDNT, the carbon *p* orbitals contribute a significant fraction of the total density of states.

**Conclusion**

Within this extensive theoretical study, we have systematically calculated the structural and electronic properties in a series of armchair and zigzag graphdiyne nanotubes via large-scale DFT calculations. Our calculations utilize the HSE06 functional (which gives accurate estimates of the bandgap compared to computationally expensive $G_0W_0$ calculations), and we present quantitative predictions of the structural relaxation energy, effective electron/hole mass, and size-scaling of the bandgap as a function of size and chirality. These calculations provide a systematic evaluation of the structural and electronic properties of the largest graphdiyne nanotubes to date (up to 1,296 atoms and 23,328 basis functions). To the best of our knowledge, a systematic study on the structural and electronic properties of GDNTs as a function of size and chirality has not been previously reported. Our calculations find that zigzag GDNTs are structurally more stable compared to armchair GDNTs of the same size. Furthermore, these large-scale calculations allow us to present simple analytical formulae to guide future experimental efforts for estimating the fundamental bandgaps of these unique nanotubes as a function of chirality and diameter. While the bandgaps for both the armchair and zigzag GDNTs can be tuned as a function of size, the conductivity in each of these two different chiralities is markedly different. Both the HOCO and LUCO in armchair GDNTs are localized on the acetylenic linkages along the circumference of the nanotube. In contrast, the HOCO and LUCO are delocalized along the entire axis of zigzag GDNTs and, therefore, both hole- and electron-transport are more facile in zigzag GDNTs compared to their armchair counterparts.

Looking forward, it would be of immense interest to understand and predict the excited-state and optoelectronic properties of these GDNTs using first-principles theoretical methods. As planar graphdiyne has garnered very recent attention as a photocatalyst for hydrogen production,[17] the use of GDNTs would offer additional electronic properties that can be tailored for these photoelectrochemical processes. For example, the bandgaps of both the armchair and zigzag GDNTs can be tuned as a function of diameter and, therefore, can be used as photo-absorbers that span a wide range of the solar spectrum. Furthermore, since both hole- and electron-transport are qualitatively different in armchair and zigzag GDNTs, these nanomaterials provide a new opportunity for modulating both charge- and energy-transfer dynamics in these photocatalytic systems.

**Acknowledgements**

We acknowledge the National Science Foundation for the use of supercomputing resources through the Extreme Science and Engineering Discovery Environment (XSEDE), Project No. TG-CHE150040.

**Table 1:** Radii, relaxation energies, binding energies per atom, effective electron mass, effective hole mass and bandgaps of armchair graphdiyne nanotubes

| Subunits | Radius (nm) | Relaxation Energy (eV) | Binding Energy per Atom (eV) | Electron Mass ($m_e$) | Hole Mass ($m_e$) | Bandgap (eV) |
|---|---|---|---|---|---|---|
| 2 | 0.29 | 8.78 | -6.9866 | 0.518 | 4.451 | 1.727 |
| 3 | 0.44 | 5.86 | -7.0543 | 0.256 | 0.412 | 1.433 |
| 4 | 0.59 | 4.41 | -7.0780 | 0.207 | 0.271 | 1.339 |
| 5 | 0.74 | 3.54 | -7.0890 | 0.187 | 0.227 | 1.300 |
| 6 | 0.89 | 2.94 | -7.0949 | 0.178 | 0.206 | 1.281 |
| 7 | 1.04 | 2.50 | -7.0986 | 0.172 | 0.195 | 1.270 |
| 8 | 1.19 | 2.19 | -7.1010 | 0.168 | 0.188 | 1.263 |
| 9 | 1.34 | 1.93 | -7.1026 | 0.166 | 0.183 | 1.258 |
| 10 | 1.49 | 1.72 | -7.1038 | 0.164 | 0.180 | 1.255 |
| 11 | 1.64 | 1.54 | -7.1047 | 0.163 | 0.177 | 1.252 |
| 12 | 1.79 | 1.39 | -7.1053 | 0.162 | 0.175 | 1.251 |
| 13 | 1.94 | 1.27 | -7.1059 | 0.161 | 0.174 | 1.249 |
| 14 | 2.09 | 1.16 | -7.1063 | 0.161 | 0.173 | 1.248 |
| 15 | 2.24 | 1.09 | -7.1066 | 0.160 | 0.172 | 1.247 |
| 16 | 2.38 | 1.01 | -7.1068 | 0.160 | 0.171 | 1.246 |
| 17 | 2.53 | 0.93 | -7.1070 | 0.160 | 0.171 | 1.245 |
| 18 | 2.68 | 0.87 | -7.1072 | 0.159 | 0.170 | 1.245 |
| 19 | 2.83 | 0.82 | -7.1074 | 0.159 | 0.170 | 1.244 |
| 20 | 2.98 | 0.76 | -7.1075 | 0.159 | 0.170 | 1.244 |
| 21 | 3.13 | 0.71 | -7.1076 | 0.159 | 0.169 | 1.243 |
| 22 | 3.28 | 0.67 | -7.1077 | 0.158 | 0.169 | 1.243 |
| 23 | 3.43 | 0.63 | -7.1078 | 0.158 | 0.169 | 1.242 |
| 24 | 3.58 | 0.60 | -7.1079 | 0.158 | 0.168 | 1.242 |
| 25 | 3.73 | 0.58 | -7.1079 | 0.158 | 0.168 | 1.242 |
| 26 | 3.88 | 0.55 | -7.1080 | 0.158 | 0.168 | 1.241 |
| 27 | 4.02 | 0.53 | -7.1080 | 0.158 | 0.168 | 1.241 |
| 28 | 4.17 | 0.50 | -7.1081 | 0.158 | 0.168 | 1.241 |
| 29 | 4.32 | 0.49 | -7.1081 | 0.158 | 0.168 | 1.241 |
| 30 | 4.47 | 0.47 | -7.1081 | 0.158 | 0.168 | 1.240 |
| 31 | 4.62 | 0.46 | -7.1082 | 0.158 | 0.167 | 1.240 |
| 32 | 4.77 | 0.45 | -7.1082 | 0.158 | 0.167 | 1.240 |
| 33 | 4.92 | 0.44 | -7.1082 | 0.157 | 0.167 | 1.240 |
| 34 | 5.07 | 0.43 | -7.1082 | 0.157 | 0.167 | 1.239 |
| 35 | 5.22 | 0.43 | -7.1082 | 0.157 | 0.167 | 1.239 |
| 36 | 5.37 | 0.43 | -7.1082 | 0.157 | 0.167 | 1.239 |

**Table 2:** Radii, relaxation energies, binding energies per atom, effective electron mass, effective hole mass and bandgaps of zigzag graphdiyne nanotubes

| Subunits | Radius (nm) | Relaxation Energy (eV) | Binding Energy per Atom (eV) | Electron Mass ($m_e$) | Hole Mass ($m_e$) | Bandgap (eV) |
|---|---|---|---|---|---|---|
| 2 | 0.51 | 2.95 | -7.0675 | 0.191 | 0.215 | 1.450 |
| 3 | 0.77 | 1.99 | -7.0902 | 0.163 | 0.171 | 1.344 |
| 4 | 1.03 | 1.49 | -7.0982 | 0.161 | 0.172 | 1.303 |
| 5 | 1.29 | 1.19 | -7.1020 | 0.169 | 0.180 | 1.285 |
| 6 | 1.55 | 0.98 | -7.1040 | 0.160 | 0.171 | 1.274 |
| 7 | 1.81 | 0.83 | -7.1053 | 0.163 | 0.174 | 1.268 |
| 8 | 2.07 | 0.72 | -7.1061 | 0.159 | 0.170 | 1.264 |
| 9 | 2.32 | 0.64 | -7.1066 | 0.160 | 0.171 | 1.261 |
| 10 | 2.58 | 0.57 | -7.1070 | 0.158 | 0.169 | 1.257 |
| 11 | 2.84 | 0.52 | -7.1073 | 0.159 | 0.170 | 1.255 |
| 12 | 3.10 | 0.47 | -7.1075 | 0.158 | 0.168 | 1.252 |
| 13 | 3.36 | 0.44 | -7.1076 | 0.158 | 0.169 | 1.250 |
| 14 | 3.61 | 0.41 | -7.1077 | 0.157 | 0.168 | 1.249 |
| 15 | 3.87 | 0.39 | -7.1078 | 0.158 | 0.168 | 1.247 |
| 16 | 4.13 | 0.37 | -7.1079 | 0.157 | 0.168 | 1.246 |
| 17 | 4.39 | 0.36 | -7.1080 | 0.157 | 0.168 | 1.245 |
| 18 | 4.65 | 0.35 | -7.1080 | 0.157 | 0.167 | 1.244 |
| 19 | 4.91 | 0.35 | -7.1081 | 0.157 | 0.168 | 1.243 |
| 20 | 5.16 | 0.35 | -7.1081 | 0.157 | 0.167 | 1.242 |
| 21 | 5.42 | 0.35 | -7.1081 | 0.157 | 0.167 | 1.242 |

**References**

1.      Iijima, S., Helical microtubules of graphitic carbon. *Nature* **1991,** *354*, 56-58.
2.      Hu, J.; Odom, T. W.; Lieber, C. M., Chemistry and Physics in One Dimension:  Synthesis and Properties of Nanowires and Nanotubes. *Accounts Chem Res* **1999,** *32*, 435-445.
3.      Zhou, X. J.; Zifer, T.; Wong, B. M.; Krafcik, K. L.; Leonard, F.; Vance, A. L., Color Detection Using Chromophore-Nanotube Hybrid Devices. *Nano Lett* **2009,** *9*, 1028-1033.
4.      Wong, B. M.; Morales, A. M., Enhanced photocurrent efficiency of a carbon nanotube p-n junction electromagnetically coupled to a photonic structure. *J Phys D Appl Phys* **2009,** *42*.
5.      Zhao, Y. C.; Huang, C. S.; Kim, M.; Wong, B. M.; Leonard, F.; Gopalan, P.; Eriksson, M. A., Functionalization of Single-Wall Carbon Nanotubes with Chromophores of Opposite Internal Dipole Orientation. *Acs Appl Mater Inter* **2013,** *5*, 9355-9361.
6.      Joo, Y.; Brady, G. J.; Shea, M. J.; Oviedo, M. B.; Kanimozhi, C.; Schmitt, S. K.; Wong, B. M.; Arnold, M. S.; Gopalan, P., Isolation of Pristine Electronics Grade Semiconducting Carbon Nanotubes by Switching the Rigidity of the Wrapping Polymer Backbone on Demand. *Acs Nano* **2015,** *9*, 10203-10213.
7.      Ford, A. C.; Shaughnessy, M.; Wong, B. M.; Kane, A. A.; Kuznetsov, O. V.; Krafcik, K. L.; Billups, W. E.; Hauge, R. H.; Leonard, F., Physical removal of metallic carbon nanotubes from nanotube network devices using a thermal and fluidic process. *Nanotechnology* **2013,** *24*.

8. Huang, C. S.; Wang, R. K.; Wong, B. M.; Mcgee, D. J.; Leonard, F.; Kim, Y. J.; Johnson, K. F.; Arnold, M. S.; Eriksson, M. A.; Gopalan, P., Spectroscopic Properties of Nanotube-Chromophore Hybrids. *Acs Nano* **2011,** *5*, 7767-7774.
9. Spinks, G. M.; Wallace, G. G.; Fifield, L. S.; Dalton, L. R.; Mazzoldi, A.; De Rossi, D.; Khayrullin, I. I.; Baughman, R. H., Pneumatic Carbon Nanotube Actuators. *Adv Mater* **2002,** *14*, 1728-1732.
10. Vohrer, U.; Kolaric, I.; Haque, M. H.; Roth, S.; Detlaff-Weglikowska, U., Carbon nanotube sheets for the use as artificial muscles. *Carbon* **2004,** *42*, 1159-1164.
11. Simmons, T. J.; Hashim, D.; Vajtai, R.; Ajayan, P. M., Large Area-Aligned Arrays from Direct Deposition of Single-Wall Carbon Nanotube Inks. *J Am Chem Soc* **2007,** *129*, 10088-10089.
12. Guldi, D. M.; Rahman, G. M. A.; Prato, M.; Jux, N.; Qin, S.; Ford, W., Single-Wall Carbon Nanotubes as Integrative Building Blocks for Solar-Energy Conversion. *Angewandte Chemie* **2005,** *117*, 2051-2054.
13. Wilder, J. W. G.; Venema, L. C.; Rinzler, A. G.; Smalley, R. E.; Dekker, C., Electronic structure of atomically resolved carbon nanotubes. *Nature* **1998,** *391*, 59-62.
14. Shao, Z.-G.; Ye, X.-S.; Yang, L.; Wang, C.-L., First-principles calculation of intrinsic carrier mobility of silicene. *J Appl Phys* **2013,** *114*, 093712.
15. Cranford, S. W.; Buehler, M. J., Selective hydrogen purification through graphdiyne under ambient temperature and pressure. *Nanoscale* **2012,** *4*, 4587-4593.
16. Jalili, S.; Houshmand, F.; Schofield, J., Study of carrier mobility of tubular and planar graphdiyne. *Applied Physics a-Materials Science & Processing* **2015,** *119*, 571-579.
17. Li, J.; Gao, X.; Liu, B.; Feng, Q.; Li, X.-B.; Huang, M.-Y.; Liu, Z.; Zhang, J.; Tung, C.-H.; Wu, L.-Z., Graphdiyne: A Metal-Free Material as Hole Transfer Layer To Fabricate Quantum Dot-Sensitized Photocathodes for Hydrogen Production. *J Am Chem Soc* **2016,** *138*, 3954-3957.
18. Jiao, Y.; Du, A. J.; Hankel, M.; Zhu, Z. H.; Rudolph, V.; Smith, S. C., Graphdiyne: a versatile nanomaterial for electronics and hydrogen purification. *Chemical Communications* **2011,** *47*, 11843-11845.
19. Li, G.; Li, Y.; Qian, X.; Liu, H.; Lin, H.; Chen, N.; Li, Y., Construction of Tubular Molecule Aggregations of Graphdiyne for Highly Efficient Field Emission. *The Journal of Physical Chemistry C* **2011,** *115*, 2611-2615.
20. Long, M.; Tang, L.; Wang, D.; Li, Y.; Shuai, Z., Electronic Structure and Carrier Mobility in Graphdiyne Sheet and Nanoribbons: Theoretical Predictions. *Acs Nano* **2011,** *5*, 2593-2600.
21. Dovesi, R.; Orlando, R.; Erba, A.; Zicovich-Wilson, C. M.; Civalleri, B.; Casassa, S.; Maschio, L.; Ferrabone, M.; De La Pierre, M.; D'Arco, P.; Noël, Y.; Causà, M.; Rérat, M.; Kirtman, B., CRYSTAL14: A program for the ab initio investigation of crystalline solids. *Int J Quantum Chem* **2014,** *114*, 1287-1317.
22. Heyd, J.; Scuseria, G. E., Efficient hybrid density functional calculations in solids: Assessment of the Heyd–Scuseria–Ernzerhof screened Coulomb hybrid functional. *The Journal of Chemical Physics* **2004,** *121*, 1187-1192.
23. Raeber, A. E.; Wong, B. M., The Importance of Short- and Long-Range Exchange on Various Excited State Properties of DNA Monomers, Stacked Complexes, and Watson–Crick Pairs. *J Chem Theory Comput* **2015,** *11*, 2199-2209.
24. Katan, C.; Savel, P.; Wong, B. M.; Roisnel, T.; Dorcet, V.; Fillaut, J. L.; Jacquemin, D., Absorption and fluorescence signatures of 1,2,3-triazole based regioisomers: challenging compounds for TD-DFT. *Phys Chem Chem Phys* **2014,** *16*, 9064-9073.
25. Foster, M. E.; Wong, B. M., Nonempirically Tuned Range-Separated DFT Accurately Predicts Both Fundamental and Excitation Gaps in DNA and RNA Nucleobases. *J Chem Theory Comput* **2012,** *8*, 2682-2687.
26. Wong, B. M.; Hsieh, T. H., Optoelectronic and Excitonic Properties of Oligoacenes: Substantial Improvements from Range-Separated Time-Dependent Density Functional Theory. *J Chem Theory Comput* **2010,** *6*, 3704-3712.
27. Wong, B. M.; Piacenza, M.; Della Sala, F., Absorption and fluorescence properties of oligothiophene biomarkers from long-range-corrected time-dependent density functional theory. *Phys Chem Chem Phys* **2009,** *11*, 4498-4508.
28. Wong, B. M.; Cordaro, J. G., Coumarin dyes for dye-sensitized solar cells: A long-range-corrected density functional study. *J Chem Phys* **2008,** *129*.
29. Allec, S. I.; Ilawe, N. V.; Wong, B. M., Unusual Bandgap Oscillations in Template-Directed π-Conjugated Porphyrin Nanotubes. *The Journal of Physical Chemistry Letters* **2016,** *7*, 2362-2367.
30. Wong, B. M.; Ye, S. H., Self-assembled cyclic oligothiophene nanotubes: Electronic properties from a dispersion-corrected hybrid functional. *Phys Rev B* **2011,** *84*.
31. Peintinger, M. F.; Oliveira, D. V.; Bredow, T., Consistent Gaussian basis sets of triple-zeta valence with polarization quality for solid-state calculations. *J Comput Chem* **2013,** *34*, 451-459.
32. Guangfu, L.; Xuemin, Q.; Huibiao, L.; Rui, Q.; Jing, Z.; Linze, L.; Zhengxiang, G.; Enge, W.; Wai-Ning, M.; Jing, L.; Yuliang, L.; Nagase, S., Quasiparticle energies and excitonic effects of the two-dimensional carbon allotrope graphdiyne: Theory and experiment. *Physical Review B (Condensed Matter and Materials Physics)* **2011,** *84*, 075439 (5 pp.)-075439 (5 pp.).